# A TAXONOMY FOR TOOLS, PROCESSES AND LANGUAGES IN AUTOMOTIVE SOFTWARE ENGINEERING

Florian Bock[1] and Daniel Homm[1] and Sebastian Siegl[2] and Reinhard German[1]

[1]Department of Computer Science 7,
Friedrich-Alexander-University, 91058 Erlangen, Germany
florian.inifau.bock@fau.de, daniel.homm@fau.de, reinhard.german@fau.de
[2]Audi AG,
85045 Ingolstadt, Germany
sebastian.siegl@audi.de

## ABSTRACT

*Within the growing domain of software engineering in the automotive sector, the number of used tools, processes, methods and languages has increased distinctly in the past years. To be able to choose proper methods for particular development use cases, factors like the intended use, key-features and possible limitations have to be evaluated. This requires a taxonomy that aids the decision making. An analysis of the main existing taxonomies revealed two major deficiencies: the lack of the automotive focus and the limitation to particular engineering method types. To face this, a graphical taxonomy is proposed based on two well-established engineering approaches and enriched with additional classification information. It provides a self-evident and -explanatory overview and comparison technique for engineering methods in the automotive domain. The taxonomy is applied to common automotive engineering methods. The resulting diagram classifies each method and enables the reader to select appropriate solutions for given project requirements.*

## 1. INTRODUCTION

Since the first definition of the term *Software Engineering* in a NATO conference report from 1968 [29], a lot of new tools, processes, programming languages and other software engineering methods have appeared. They provide different key-features, advantages and disadvantages and they especially differ in their associated application domain. Within these different domains, the automotive sector is the focus of this paper.

Cars have developed from being completely mechanical in the early 20th century to being electromechanical in the subsequent decades until finally reaching the present-day's complexity in terms of hardware and software. Especially in case of software development, such aspects like the quantity of functions embedded in the car or the binary code size have increased exponentially [10],[11],[13]. To face these challenges, on the one hand, the hardware is continuously improved by more powerful components. On the other hand, the high climax in software challenges cannot be solved just by hardware improvements, but requires evolution in software engineering. The required efforts can be divided into two categories: runtime efforts and design efforts. Runtime efforts are concerned with the optimal execution of complex code on the hardware. Here, software engineering improvements are hardly feasible. Hence, this is not in the focus of this paper. Design efforts relate to the efficient specification of complex software, which results in a need for good software engineering methods. This is the key topic of this paper.

The development cycle for a car series was reduced by about 25% during the past decades [33], while the development complexity increased. Using the same well-established engineering methods would result in a great demand for new man-power, which is not economical. Resources have to be ideally utilized. New software engineering methods can help to reach this goal. However, new methods often differ in several aspects and hence, for each scenario in the development process, different adequate methods are available. To be able to choose the proper approach for a given project scenario, the common methods placed on the market have to be examined, classified and compared to offer this information and classification to potential users. Especially the comparison of methods of fundamentally different types, for example processes and tools, may seem like trying to compare apples and oranges, due to the largely mismatching set of characteristics. Common comparison techniques are not applicable, because they require measurable, quantifiable and matchable characteristics to work properly. Nevertheless, a comparison by any means is necessary to be able to come to a decision for a suitable method in a specific project scenario. Therefore, we introduce a taxonomy, which allows such a classification and is tailored to the automotive domain. We applied it to the main methods available in this area. Thus, a compact and comprehensible overview of the current market situation is also given.

We conducted a survey among 15 representatives from different companies and departments to verify the assumptions established in this paper. It consists of 15 questions. The raw survey data and the survey form can be viewed online [9]. Two-thirds of the respondents work for a car manufacturer, one-fifth in research and the rest for automotive suppliers. Their areas of activity consist of requirements engineering, system architecture, implementation, test, documentation, change-management, administration/organization and miscellaneous topics with an emphasis on requirements engineering and test. The self-evaluation of the respondents regarding their software-engineering skills revealed an overall high average skill level. 47% are decision makers. The age of the respondents ranges from 20 to 49.

## 2. TERMINOLOGY

To be able to describe the classification scheme outlined in this paper, several basic terms have to be taken into account: *Tool*, *Method*, *Process*, *Language* and particularly *General Programming Language* and *Domain Specific Language* [6],[24],[32]. The terms already allow a three-part classification of software engineering approaches into: *Tool* as a piece of software, *Process* as a general description of a procedure, and *Language* as a well-defined mode of communication or specification. The term *Method* is applicable to all of them, because it is a general description of a procedure, which is implied as well in tools, processes and languages.

The classification of available market solutions into this pattern is not always distinct and might require a deep analysis of each approach. There is the possibility that some methods may fit in more than one category.

Terms and subcategories of languages are difficult to determine and apply, because they are partly used quite different depending on the domain or user group. For instance according to [26], languages can be subdivided into *GPLs* and *DSLs*, whereat in [35], *programming-* and *modeling-languages* are employed. As a compromise, the categorization displayed in figure 1 is used below at which *Others* stands for natural languages (e.g. *English*) without any programmatic background.

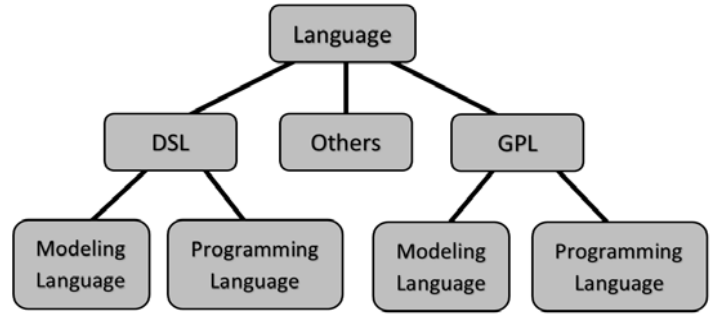

Figure 1. Classification of Languages

The correct classification is not as clear as it might appear at first glance. The main differentiator is obviously the limitation of *DSLs* to a specific domain, whereas *GPLs* can be applied to all domains. Indeed, this is only sufficient as sole distinction feature for some candidates e.g. *C++*, which is clearly a *GPL*. Other languages like the *Unified Modeling Language (UML)* [31] or the *System Modeling Language (SysML)* [30] are apparently limited to a specific domain, but are categorized as a *GPL* [30]. Hence, a more detailed distinction method is required, which can partly be derived from [35] and [26]. This classification task is succinctly described in chapter 4.

## 3. RELATED WORK

There already exist various taxonomies that help to classify software engineering methods. To the best knowledge of the authors, the main approaches have been selected and are elaborated in detail below, with special focus on the applicability to the automotive domain and its requirements.

Blum [8] proposed a classification scheme for engineering methods that distinguishes between *Problem-oriented* and *Product-oriented* attempts as well as between *Conceptual* and *Formal* ones. A matrix of these two differentiation schemes allows a simple classification. However, it does not take into account topics like engineering steps, modeling roles or the automotive context.

Kitchenham [25] focused on the DESMET evaluation method. She identified evaluation types that enable a comparison between different software engineering methods and tools: Quantitative types, qualitative types and other types. The evaluation types are empirical attempts. They need a large amount of data about an engineering method to allow a categorization. This data can only be obtained for well-defined and ready-to-use methods, which can be tested in real projects. Attempts without any existing application or with limited data are not covered by this taxonomy.

Babar et. al. [5] introduced a taxonomy to compare general software architecture analysis approaches. It consists of 17 evaluation questions grouped in the four categories context, stakeholders, contents and reliability. Their taxonomy is limited to software architecture analysis methods. Additionally, the automotive context is missing.

Hofmeister et. al. [23] proposed a taxonomy for architectural design methods that provides two kinds of comparison techniques: activity-based and artifact-based. The former involves an architectural analysis, synthesis and evaluation, whereas the latter considers architectural concerns or candidate architectural solutions. The taxonomy is lacking the automotive focus.

Broy et. al. [14] defined a taxonomy for engineering tools in the automotive domain. It classifies tools by vertical domain-related and horizontal domain-independent aspects. The former considers language aspects whereas the latter concerns aspects of the tool framework. Prior to the classification of a tool, empirical data has to be obtained by investigating its toolbars/menu items and identifying the underlying functionality as domain-related or -independent. The taxonomy is focused on the automotive domain, however, the limitation to tools excludes languages and processes without an integrated tool.

The main deficiencies of the above summarized approaches are:

- The lack of an automotive focus. Therefore, the results cannot be applied directly to that domain.

- The limitation to a particular type of engineering method. Methods of different types cannot be compared.

- The primarily use of quantifiable characteristics to compare methods. Such approaches are benchmarks with the objective of providing a method ranking. This requires the collection of much data for each method and is only applicable for methods of the same type.

Such limitations are, as already described in the introduction, not feasible in some project settings. Especially at the project start, diverse methods, tools and processes with their individual characteristics are candidates and therefore under investigation. A comparison cannot be accomplished by the above reviewed taxonomies. Hence, a new comparison technique is required, which is developed in this paper as new, generally applicable and lightweight taxonomy for the automotive domain. Its main aim is to guide the decision making by the use of an appropriate overview of the methods in question.

## 4. TAXONOMY FOR THE AUTOMOTIVE DOMAIN

Two-thirds of our survey respondents are not satisfied with the methods currently used in their environment. Their willingness to introduce new approaches into their established workflows is quite high. A suitable and lightweight taxonomy that fits to the automotive domain and provides an overview of available methods helps to improve the situation. It may also increase the willingness of the department to introduce new methods, which is low according to our survey respondents. Such a taxonomy has to be plausible, adaptable to methods of varying types, and clear. It can basically be visualized textually or graphically. If the methods, which should be compared, share the same type and are directly comparable as to e.g. their key features, a textual or tabular approach might be adequate. In the given context, this is obviously not the case. Therefore and by reasons of simplicity and clarity, a graphical taxonomy seems to be the most appropriate way to offer an easy and understandable decision pattern for a wide range of different engineering methods. Primary goal is not to evaluate the performance of the methods and create a ranking, but to offer a lightweight, comprehensible and clear overview and comparison pattern.

As most of the methods commonly used in the automotive domain are based on the *V-Model* [12], it can be taken as a reliable base model. This is also verified by our survey, in which all of the respondents indicate familiarity with it [9]. Though, it is rather generic and therefore neither limited to a specific domain, nor enriched with automotive terms and views. As a result, the automotive context is considered by using a level model that represents the different modelling steps during software development in the automotive domain. Instead of proposing a completely new level model, an already specified and field-tested one is used: the model incorporated in the

*EAST-ADL* approach [18] (cf. chapter 5.3). This ensures both adaptability and applicability for the given context.

The level model from the *EAST-ADL*-specification consists of four consecutive abstraction levels [18]:

- *Vehicle Level*: A solution-independent, abstract description of the target car functions (e.g. driver assistance systems). This includes use cases, requirements and high-level descriptions of features- and functions, all of them as graphical as well as textual artifacts.

- *Analysis Level*: A functional black-box decomposition with interface information. The artifacts from the level above are enriched with additional information. The resulting system is designed as a black-box architecture, consisting of several blocks with raw specifications about their interactions, e.g. which information should be collected from outside the system and which output should be returned.

- *Design Level*: A functional white-box decomposition with hardware information, e.g. the type of controller or sensor used in the target system. The black-box specification is filled with the inner behavior in the form of abstract algorithms, state machines and additional information. Thus, a complete system behavior model is created.

- *Implementation Level*: An implementation of the car functions. Here, the system model created in the previous levels is implemented in the target language and delivered to the target platform (for example a controller or another embedded device). The initially defined car functions are practically usable and testable.

Each of the levels contains both specification and test of the particular artifacts.

These levels with their descriptions resemble the phases of the *V-Model*. Hence, the phases and the levels can be overlaid (cf. figure 2). This is valid, because the layered architecture from *EAST-ADL* is derived from the *V-Model* [7].

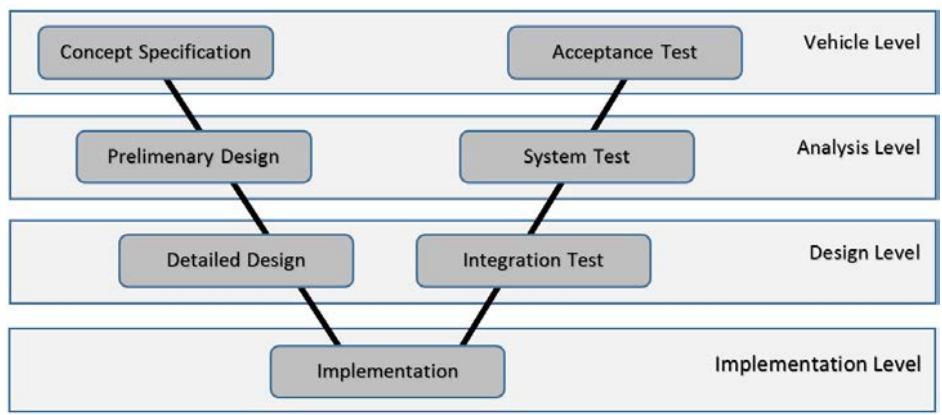

Figure 2. Overlay of the *V-Model* and the *EAST-ADL*-levels

In addition, the type of an engineering method should be reflected in the diagram. As already described in chapter 2, the terms *process*, *tool* and *language* are applicable, whereupon *language* can be subdivided in *DSLs* and *GPLs*. Due to the fact that some methods cover more than one level or step of the *V-Model*, it is not sufficient to simply note a method textually in the diagram. The use of formatted bars as graphical representation for the different methods and their coverage of software development steps seems appropriate.

The lines and the color (in this case gray-scale) of a simple bar are modified in a readable and constructive way to encode the categorization information as combination of *language*, *process* and *tool* (cf. figure 3). This formatting rules ensure that the diagram stays simple and readable.

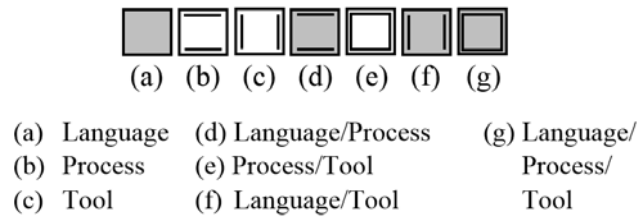

(a) Language  (d) Language/Process  (g) Language/
(b) Process   (e) Process/Tool          Process/
(c) Tool      (f) Language/Tool         Tool

Figure 3. Encoding for engineering methods

Additionally, the general type of language should be included in the notation. The background is altered to reflect this information*: dark-gray* for *DSLs* and *light-gray* for *GPLs*. To determine the language type, the classification patterns from [26] and the information from the respective language provider are used.

## 5. EVALUATION

There are several software engineering methods currently available on the market. This paper focuses on the most common and established ones: *Rational Harmony*, *AUTOSAR*, *EAST-ADL*, *MATLAB/Simulink/TargetLink*, *SCADE*, *ADTF*, *RUP/EUP* and *SimTAny*. We applied our taxonomy to these approaches, which yields their classification depicted in figure 4a/4b. Due to clarity reasons, the phases of the *V-Model* are abbreviated (cf. figure 2) and the approaches are spread across two diagrams.

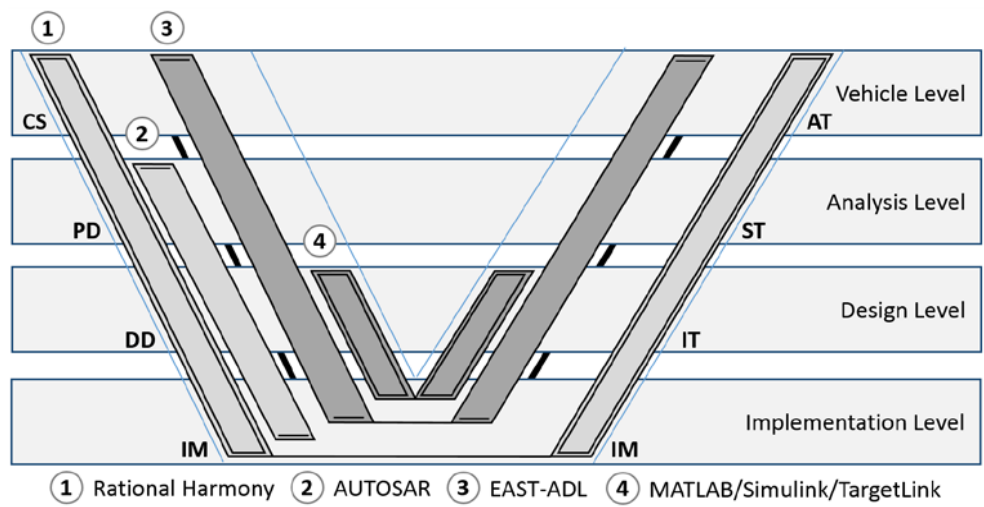

Figure 4a. Automotive specific taxonomy applied to common engineering methods

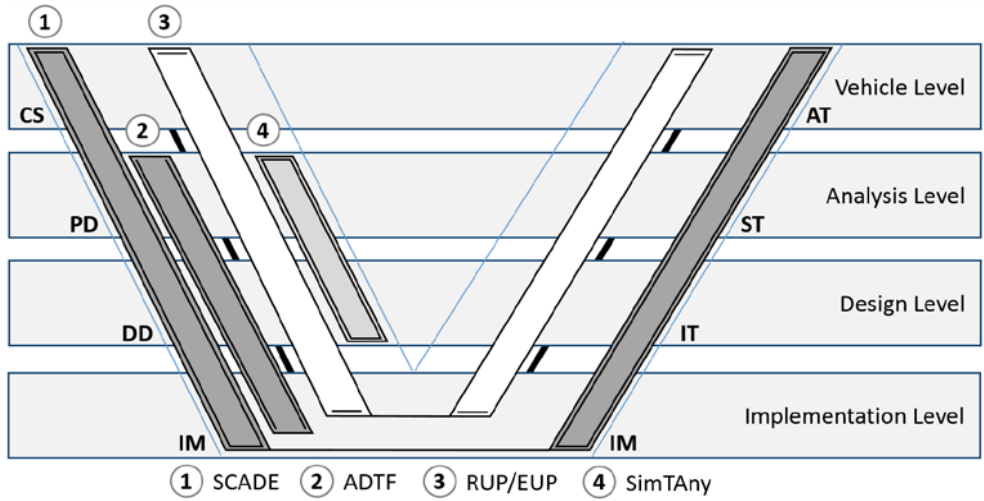

Figure 4b. Automotive specific taxonomy applied to common engineering methods

The diagram can be used to determine an appropriate solution for a given development scenario and to exclude methods, that do not fulfil the project requirements. As depicted in our survey, the knowledge of individual persons and departments about the characteristics of a specific method, its availability or even its existence varies considerably [9]. Our taxonomy deals with this fact by providing an overview with comprehensible information, which can be used without the need for extensive knowledge of each method. This overview also contains the information, whether a tool aspect is included in the method or not. This can be crucial for a reliable decision.

### 5.1. Rational Harmony

*IBM Rational Harmony* [22] is an iterative software modelling process based on the *V-Model* [12]. It is split into two sequenced sections (cf. figure 5). First, the system behavior is modeled as *SysML* model with regard to requirements and use cases. The second step enhances this model and transforms it into an *UML* model, which contains all information necessary to generate both the required system artifacts and the target code. The simulation of the created models and different validation/verification methods are also part of the process and tooling. To increase the usability, semi-automatic wizards assist with the different modelling steps.

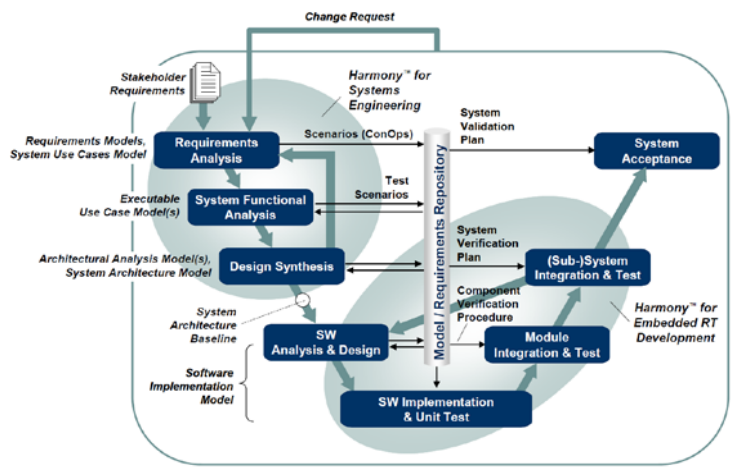

Figure 5. *Harmony* Process Overview [22]

*Rational Harmony* is designed as process with different steps and covers all phases/levels from the taxonomy. As sole implied languages, *UML* and *SysML* are used, which are classified as *GPL*. *Rational Harmony* is always delivered within the tool *Rational Rhapsody*.

Even though being available since 2006 [22], *Rational Harmony* was introduced quite recently in the automotive domain. The implied process steps are generally applicable, so they can easily be adopted for the specific requirements of the domain. Nevertheless, it is not yet widely deployed at present, which is reflected by our survey. Only one-fifth of the respondents indicate the use of *Rational Harmony* in their departments [9].

### 5.2. AUTOSAR

The *AUTomotive Open System ARchitecture (AUTOSAR)* [2],[21] is a software architecture standard widely used in the automotive domain and developed by the *AUTOSAR* development partnership. Its focus is the implementation and realization of automotive software systems. To abstract and standardize the development, a layered software architecture is used (cf. figure 6). When utilizing *AUTOSAR*, all required software artifacts for the target car function are located at the *Application Layer*. They consist of so-called *Software Components (SWCs)*, which enfold both the algorithms (which are enclosed in *Runnables*) and the wrapper-code for the car function. To simplify the exchange of model artifacts, a well-defined XML-scheme is used to store all information.

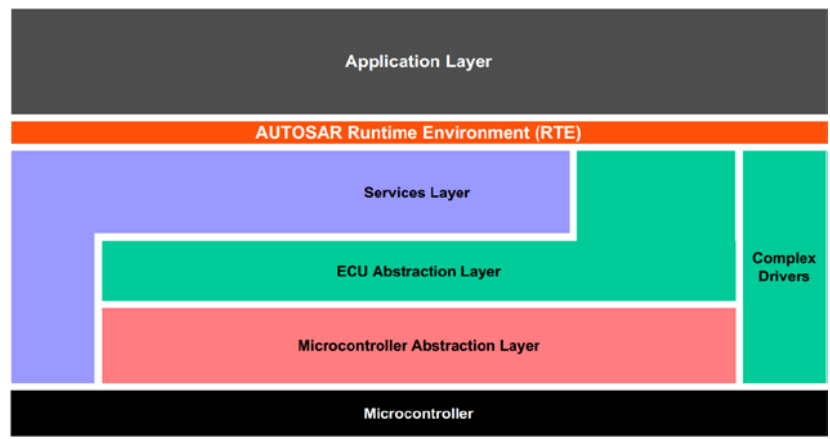

Figure 6. The *AUTOSAR* layered architecture [3]

The software architecture models contain abstract as well as low-level information, so the adoption starts within the *Analysis Level* and lasts until the *Implementation Level*. Tests or test strategies are not specified. *AUTOSAR* includes specifications, but no implementation by itself is implied, although external tools exist. Lines of action, which form a process, are provided and *GPL* aspects are available in terms of model definitions. There is, by default, no *DSL* embedded, but an external add-on exists (*ARText* [4]). It is a language framework to build user-defined *DSLs* for *AUTOSAR*.

*AUTOSAR* was initially developed and designed 2005 to be used in the automotive context and is already wide spread in the domain. Our survey shows, that 87% of the respondents are familiar with *AUTOSAR* and 60% already work with it [9].

## 5.3. EAST-ADL

The *Electronics Architecture and Software Technology - Architecture Description Language (EAST-ADL)* [7],[18] is developed and enhanced by the *EAST-ADL Association*. It uses *AUTOSAR* and additionally covers aspects like non-functional requirements, vehicle features and functional/hardware architecture details. The models are categorized in four different abstraction levels (cf. figure 7 and chapter 4). The process starts with a rough initial vehicle model that is enriched during the development, until it is in a highly detailed state and realized as *AUTOSAR* model.

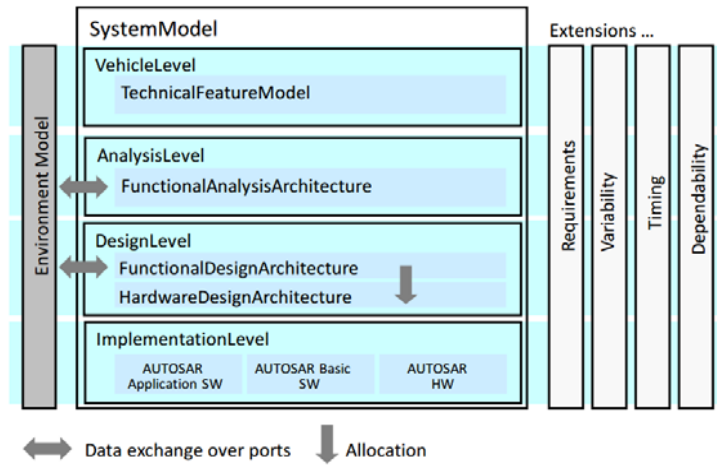

Figure 7. The *EAST-ADL* abstraction layers [7]

*EAST-ADL* covers all phases of our taxonomy, as we make use of its level system. Nevertheless, it is discussable, whether the implementation level is also covered, because it embeds *AUTOSAR* instead of a distinct solution. In the view of the authors, this can be ignored, because many methods imply predefined languages (which omits extensive redevelopment). *EAST-ADL* includes lines of action for the implementation level. This can be seen as part of a major surrounding process. Other phases are performed by use of own languages or language aspects, which can be acknowledged as *DSL*. A tool is not comprised, though some implementations are available.

*EAST-ADL* contains many information directly related to the automotive domain. The integration and use is therefore easy, whereas the lack of a proper implementation or tool for many years prevented the distribution in the domain. In line with this, none of the respondents of our survey uses *EAST-ADL* and it is scarcely known [9].

## 5.4. MATLAB/Simulink/TargetLink

*MathWorks MATLAB/Simulink* and *dSPACE TargetLink* [17],[28] compose a software modeling framework used to create a software model and its derived target code. Simulink is a graphical data flow modeling language embedded in the *MATLAB* computing environment. Models created in *Simulink* consist of so-called *blocks* (functional entities), which can be linked to each other and are taken out of a predefined block library. The models are closely related to the hardware structure, which also becomes apparent in the type of blocks available in the library, e.g. bus-, mux-/demux or gain-blocks. *TargetLink* provides target source code generation out of the created models. Testing, verification and validation methods are also available.

The method is started at the *Detailed Design* phase and continues until the corresponding *Integration Test*. MATLAB is the basic tool framework. It is mandatory for the use of *Simulink*, a graphical *DSL* used to create the required models. *TargetLink* is used to create target source code out of the models. No lines of action are included.

The *MATLAB/Simulink/TargetLink*-tool chain is one of the major software engineering frameworks currently used in the automotive domain. This is also illustrated in our survey, where at least two-third of the respondents already use the tool chain and more than 86% are familiar with it. However, it mainly lacks possibilities to design the system architecture or to include requirements at an abstract level. Consequently, the system engineering in this case is rather bottom-up and implementation-related instead of being top-down and iterative as required by the *V-Model*.

### 5.5. SCADE

The *Esterel Safety Critical Application Design Environment (SCADE)* [20] is a software development framework initially grounded in the avionics industry. It consists of four different tools, whereof *SCADE Suite* is focused on model-based software development. As basis, the formal, synchronous and data flow-oriented *DSL Lustre* [15] is used, which utilizes graphical models to describe the system-underdevelopment. The *SCADE Suite* includes methods for validation/verification and code generation.

The *SCADE*-tool chain covers the complete *V-Model*, so all phases of the taxonomy are included. With *Lustre*, a *DSL* is used. *SCADE* contains detailed process information and lines of action.

*SCADE* is well-known in its initial application area, the avionic industry, but has recently been introduced to first automotive projects. Still, an adoption for this new context requires a certain amount of modifications, e.g. the introduction of automotive-related concepts and definitions. In our survey, none of the respondents practically uses *SCADE*, whereas one-fifth are at least aware of this method.

### 5.6. ADTF

*Automotive Data and Time-Triggered Framework (ADTF)* [19] is a software modeling framework aiming at the development of driver-assistance features. *ADTF* allows real-time data playback and provides visualization features that are used to simulate the created models and evaluate it according to defined timing constraints. This guarantees, that both the simulation on the development system and on the target system act and react similar. The *ADTF*-models consist of graphical representations of functions, so-called *filters*, with their inputs and outputs (e.g. signals). As data source, different standardized sources like CAN or camera data can be used simultaneously and synchronized.

*ADTF* is a tool with focus on the development of car functions. It ranges from the *Analysis Level* until the *Implementation Level* with the integration of production code. Testing is limited to simple manual tests. Lines of action are not included, whereas the models are created with help of a graphical *DSL*. The functional range lacks detailed architecture and testing features.

*ADTF* was initially developed for the automotive domain in Germany in 2011. This, in conjunction with our survey being carried out in the environment of German car manufacturers and their suppliers explain the high familiarity of the respondents with *ADTF* and the utilization rate of 50% [9]. In foreign markets, this rate would be much lower. Hence, the use of *ADTF* is limited so far to German car manufacturers.

## 5.7. RUP/EUP

The *IBM Rational Unified Process (RUP)* [1],[27] is an iterative software development process. It is split into four phases that handle the project definition, system architecture, implementation and delivery. Each phase contains a set of engineering disciplines, which may occur iteratively. Beside the general process model, *RUP* contains best practices, templates and checklists to support the developer. The complete process setup and the importance of each discipline for each particular phase is shown in figure 8, at which the ordinate indicates the required time and effort at a specific time.

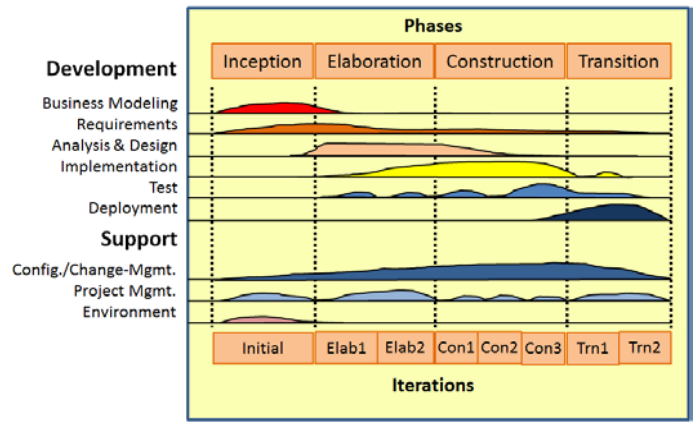

Figure 8. The *RUP* phases and disciplines [1],[27]

An enhancement to *RUP* is proposed as *Enterprise Unified Process (EUP)* [1]. It adds two new phases, that handle maintenance and retirement. Additionally, new disciplines are added (cf. figure 9). The intention is to cover the more generic and development-independent topics like personnel administration.

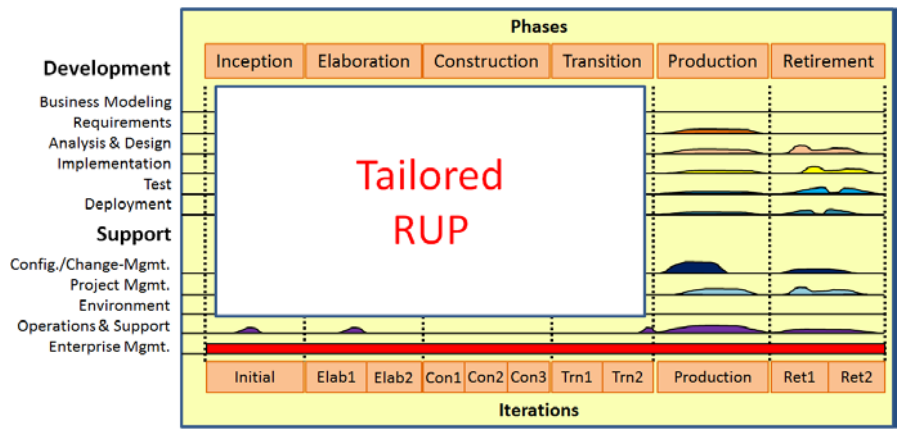

Figure 9. The *EUP* phases and disciplines [1],[27]

The development ranges from the specification to the retirement of the finished product, so all levels of the taxonomy are covered. According to [1], both methods are processes with no integrated languages or tools. To make use of them, a separate implementation is required which is not part of the original definition. Anyway, work flows and process steps can be adopted for given project scenarios.

Both *RUP* and *EUP* are primarily general processes without an implementation or any automotive focus, so the practical use in the automotive context is rather limited. Our survey states, that none of the respondents actually uses *RUP/EUP* and only a minor part is familiar with them [9].

### 5.8. SimTAny

*Simulation and Test of Anything (SimTAny)* [16] (formerly known as *VeriTAS*) represents a framework that provides the *test-driven agile simulation* (*TAS*) process and a respective tool chain. The process specifies, that the system and the usage model are derived separately from the requirements and specified by individual *UML* models (see figure 10). A respective simulation model is automatically generated from the system model and test cases are automatically derived from the usage model. Subsequently, the simulation model is run together with test cases in a simulation. An implementation of the system or the hardware is not required. Thus, it is possible to identify modeling errors or inconsistencies in the system model and/or the usage model and validate them early in the development process.

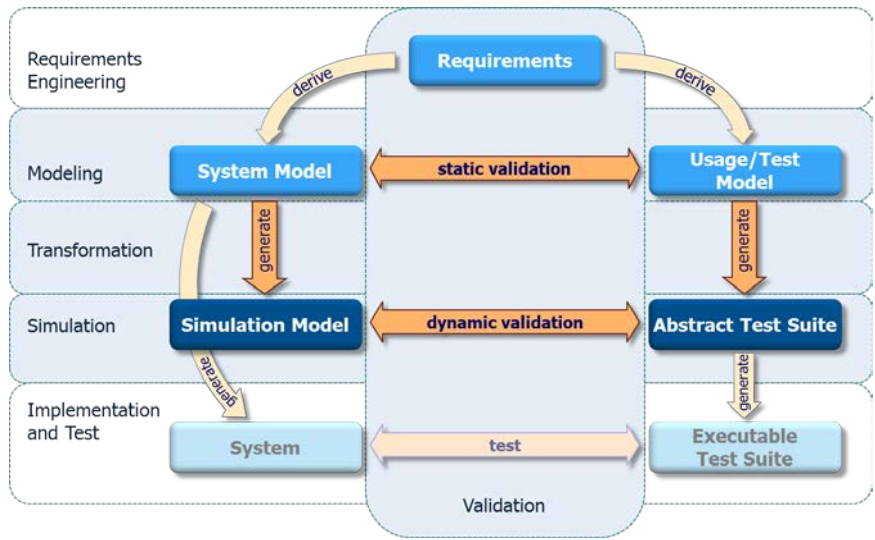

Figure 10. Test-driven agile simulation process provided by *SimTAny* [34]

Although the integration of requirements is possible, it is not the focus of *SimTAny*. Therefore the Analysis Level is the actual starting point. *SimTAny* places emphasis on the simulation of the system to be developed and does not include production code. So the *Implementation Level* and all subsequent levels are not covered. A surrounding process and a method implementation are part of *SimTAny*. As specification language, *UML* as single *GPL* is used.

*SimTAny* is mainly applied by academics or in research and therefore not used in the automotive domain so far. First projects to introduce it to the domain are currently running. Unsurprisingly, in our survey, only respondents located in research already work with *SimTAny* [9].

### 6. CONCLUSION AND FUTURE WORK

Selecting a software engineering method, that satisfies the requirements of an automotive project, is a difficult task. In order to aid the decision making, a well-structured overview, as well as a possibility to compare the features of the available approaches are required. There exist several taxonomies that provide such an overview, however, they mainly lack the automotive focus or are restricted to a specific software engineering method type. As outlined in the introduction, such an overview can be necessary for a development decision in a given project scenario, even if the investigated methods differ extensively.

That is why this paper outlines a new taxonomy for software engineering approaches focused on the automotive domain. It consists of a combination of the general *V-Model*, the level model taken out of *EAST-ADL* and the enrichment with the indication, whether *GPLs* or *DLSs* are included. Due to clarity and simplicity reasons, the results are depicted in a diagram (cf. figure 4). This allows the reader to easily compare several possibly quite different engineering approaches.

The introduced taxonomy has been applied to currently established key-methods in the domain. The result is a well-structured overview that serves as a compendium and exemplifies the approach. This approach has been reviewed in our survey by the respondents to get an indicator, how helpful, self-explanatory and useful this taxonomy appears to the target user group. The resulting evaluation values e = 6.36 with e ∈ [1, 9], 1 as representative for *not helpful* and 9 for *helpful*. This is sufficient to state the taxonomy as helpful, though this value can be increased by adding more information to the taxonomy or applying it to more different methods to provide a diversified information base for project decisions.

The taxonomy approach described in this paper is the first step in the development of a detailed classification pattern for software engineering methods in the automotive domain. The proposed format and diagram can be prospectively enriched with more classification information or can be extended with new phases/levels. As depicted in our survey, there are several additional characteristics of engineering methods that are more or less important for engineers [9]:

- important: support, extensibility, documentation, training courses

- neutral: amount of features, market share, price

- unimportant: familiarity of the manufacturer

These values cannot be linked with all types of engineering methods, e.g. processes partly have no manufacturers. Instead they are defined by standardization organizations. As a result, this list of characteristics is not yet included in our taxonomy. There are two ways of incorporating these values into the decision process. First, the values can be included by taking a subset of characteristics, that is matchable to the investigated methods and enriching the taxonomy with this subset. Second, our taxonomy can be used to constrain the list of investigated methods and afterwards, other taxonomies (e.g. developed by Broy [14]) can be used in combination with the whole set of characteristics to determine a final solution for the given project scenario. In both cases, our taxonomy serves as first easy-to-use decision guidance.